\documentclass[twoside,english,nofootinbib,preprintnumbers,aps,11pt,showkeys]{revtex4}

\usepackage{euscript,amssymb}
\usepackage{amsthm}
\usepackage{graphicx}
\usepackage{amsfonts}
\usepackage{amsmath}
\usepackage{amssymb}
\usepackage{fancyhdr}
\usepackage{epsfig}
\usepackage{color}
\usepackage{esint}
\usepackage{soul}
\usepackage{afterpage}
\usepackage[papersize={8.5in,11in}]{geometry}

\begin{document}

\title{When Isolated Horizons met Near Horizon Geometries}

\author{Jerzy Lewandowski}
\email[]{Jerzy.Lewandowski@fuw.edu.pl}
\author{Adam Szereszewski}
\email[]{Adam.Szereszewski@fuw.edu.pl}
\author{Piotr Waluk}
\email[]{Piotr.Waluk@fuw.edu.pl}
 
\affiliation{\vspace{6pt} Instytut Fizyki Teoretycznej, Uniwersytet Warszawski, ul. Pasteura 5, 02-093 Warsaw, Poland}
\begin{abstract}
There are two mathematical relativity frameworks generalizing the black hole theory: the~theory
of  isolated horizons (IH) and the theory of  near horizon geometries (NHG). We~outline here and discuss
the derivation of the NHG from the theory of IH by composing  spacetimes from IH. The simplest but still quite general  class of solutions to Einstein's  equations of this type defines spacetimes  foliated by Killing horizons emanating from  \mbox{extremal} horizons.  That derivation, clearly being a link between the two frameworks, seems to be unknown to the
NHG researchers and is hardly acknowledged in  reviews on the IH. This lecture was  a contribution to the Mathematical
Structures session of the 2nd LeCosPA \linebreak International Symposium ``Everything about Gravity'' celebrating the centenary of Einstein's General Relativity on December 14--18,  2015 in Taipei.
\end{abstract}


\maketitle

\section{Introduction}
A family of solutions to 4d-Einstein's equations was constructed in \cite{PLJ} from   the
following data:   
\begin{itemize}
\item $S$ --- a 2 dimensional manifold diffeomorphic to a sphere,
\item  $g=g_{AB}dx^Adx^B$ --- a metric tensor on $S$,
\item $\omega\ =\ \omega_Ad x^A  $ --- a differential 1-form on $S$,
\end{itemize}
such that  the following (tensor) equation is satisfied:
\begin{equation}
\label{theequation}
D_{(A}\omega_{B)}\ +\ \omega_A\omega_B\ -\ \frac{1}{2}R_{AB}\ =\ 0,
\end{equation}
where $D_A$ and $R_{AB}$ are, respectively, the torsion free covariant derivative and the Ricci tensor, defined on $S$ by the metric $g_{AB}$.  

The corresponding  solution to the vacuum Einstein equations  is a metric tensor $g_{\mu\nu}$ defined on 
	\[ S\times \mathbb{R}\times\mathbb{R} \] 
in terms of coordinates \[ (x^\mu)=(x^A,v,u), \] 
in the following way
\begin{equation}
\label{themetric} g_{\mu\nu}dx^\mu dx^\nu\ \ :=\ \ g_{AB}dx^Adx^B\ -\ 2du\left(dv -2v\omega - \frac{1}{2}v^2(D_A\omega^A  + 2\omega^A\omega_A)du\right)   
\end{equation} 
%

\section{Foliation by Killing horizons}

What is special about each of the solutions (\ref{themetric}) is that the corresponding spacetime is foliated by Killing horizons.
They are the surfaces 
\begin{equation}
\label{u=const} u\ =\ {\rm const} 
\end{equation}
of the topology $S\times \mathbb{R}$ and 
\begin{equation}
v\ =\ 0 .
\end{equation}
The Killing vector corresponding to each value $u=u_0$ is 
\begin{equation}
K^{(u_0)}\ =\ v\partial_v - (u-u_0)\partial_u .
\end{equation} 
This was exactly the aim of performing the construction in \cite{PLJ}: to compose a spacetime out of horizons.     

An unexpected property, though,  was  the emergence of   an extremal   \mbox{Killing horizon}
\begin{equation}
\label{v=0}  v\ =\ 0
\end{equation}
of the Killing vector 
\begin{equation}
K\ =\ \partial_u.
\end{equation}
All the horizons (\ref{u=const}) foliating the spacetime emanate from that extremal horizon.  


The 1-form $\omega_Adx^A$,  considered on a spacial slice 
	\[ v\ =\ {\rm const} \]
of any of the  horizons  (\ref{u=const}), coincides with  minus the rotation 1-form potential,~namely 
\begin{equation}
\nabla_A K^{(u_0)}\ =\ -\omega_A K^{(u_0)} .
\end{equation} 
Considered on each spacial slice 
\[  u\ =\ {\rm const} \] 
of the extremal horizon  (\ref{v=0}), on the other hand, the 1-form $\omega_Adx^A$  coincides 
exactly with  its rotation 1-form potential, that is,
the pullback of the spacetime covariant derivative $\nabla_\mu K$ to the slice
is
\begin{equation}
\nabla_A{K}\ =\ \omega_A K .
\end{equation} 
       
The equation (\ref{theequation})  coincides with the vacuum constraint that has to be satisfied by every induced metric tensor $g_{AB}$ and rotation 1-form potential $\omega_A$ on an extremal Killing or, more generally, isolated horizon,
at which the vacuum Einstein equations hold \cite{ABL}.  The equation was studied in \cite{LPuniq} for compact 
2d-surfaces.  It was proven therein that 
the only axisymmetric  solutions $(g,\omega)$ admitted by $S$ diffeomeomorphic to $S^2$ are those provided by the horizons of extremal Kerr solutions. This may be interpreted as a quasi-local  version of the black hole uniqueness theorem.  Those solutions correspond to the Horowitz 
limit \cite{Horowitz} of the extremal Kerr spacetimes, also known as the Kundt class solutions of the 
Petrov type D \cite{ExactSolutions}. The equation (\ref{theequation}) was farther studied in  \cite{Jez,JezKerr}.  However, intriguingly enough,  still very little is known about a generic, axially non-symmetric case.   It is not even known  whether there exists any non-axisymmetric solution to eq.~(\ref{theequation}) on $S$ diffeomorphic to \mbox{2-sphere.}  
The other compact 2d-$S$ cases are either excluded --- the higher genus case --- or trivialized by the equation (\ref{theequation}): on $S$  diffeomorphic to 2-torus, the only solutions are flat $g$ and \mbox{$\omega=0$.}
It might be possible, though, that inclusion of a non-zero cosmological constant in the calculations would loosen these restrictions somehow. 
More specifically, as the argument relies on the  definiteness of
$\omega_A\omega^A -\Lambda$. A negative cosmological constant yet strengthens that
property, and even eliminates the 2-torus case,  allowing only the 2-sphere. However, a positive cosmological constant destroys the positive definiteness
and we have no  argument against the higher genus.

Above,  Einstein's  vacuum equations can be replaced by the Einstein--Maxwell vacuum equations. The extremal horizon constraint (\ref{theequation}) is then suitably generalized by the presence of the Maxwell field. The uniqueness theorem still holds for the resulting extremal horizon constraint equation, with the Kerr solutions replaced by the 
Kerr--Newman solution \cite{LPuniq}. The suitable generalization of the  metrics (\ref{themetric}) is also available \cite{PLJ}.   Those metrics still have the form  
\begin{equation}
\label{NHmetric}
g_{\mu\nu}(x,u,v)dx^\mu dx^\nu\ =\ g_{AB}(x)dx^Adx^B\ -\ 2du\left(dv -2v\omega_A(x)dx^A - 
\frac{1}{2}v^2H(x)du\right),
\end{equation}
where the function $H(x)$ depends now on both $\omega_A(x)$ and the Maxwell field \cite{PLJ}.

\section{Near horizon limit}

Yet more generally, metric tensors of the  form (\ref{NHmetric}), with arbitrary $g_{AB}(x)$, $\omega_A(x)$, and $H(x)$ are
 called  near horizon geometries \cite{LivRevNHG}. 
The vacuum  Einstein equations imposed on such a metric are equivalent to  (\ref{theequation},\ref{themetric}). In the case with matter,  the equation (\ref{theequation}) is  generalized to
\begin{equation}
\label{theequationmatt}
D_{(A}\omega_{B)}\ +\ \omega_A\omega_B\ -\ \frac{1}{2}R_{AB}\ =\  \frac{1}{2}{\cal R}_{AB},
\end{equation}
where the new term ${\cal R}_{AB}$ is the pullback of the spacetime Ricci tensor to a spacial slice of an extremal Killing horizon \cite{LPhigh}. The corresponding change in (\ref{themetric})  amounts to a suitable
modification of the function $H(x)$  \cite{LivRevNHG}.   

The class  of  metric tensors  (\ref{NHmetric})  was  derived by a  neat argument \cite{Reall} from
the general class of metric tensors admitting an extremal Killing horizon. Let us recall. Suppose a spacetime 
metric $g$  admits a Killing horizon. Locally, in the neighbourhood of the horizon, $g$  can be written as
\begin{equation}
\label{ghoriz} g\ =\  g_{AB}(v,x)dx^Adx^B - 2du\left(dv -2v\omega_A(v,x)dx^A + (v\kappa H_0(v,x)+\frac{1}{2}v^2H(v,x))du\right).  
\end{equation}     
The Killing vector is 
	\[ K\ =\ \partial_u \]
and the Killing horizon is the surface
	\[  v\ =\ 0. \]
To remove the ambiguity of $\kappa$, we demand that $H_0\big|_{v=0}\equiv -1$. Apart from that, $g_{AB}$, $\omega_A$, $H_0$, and $H$ are \emph{a priori\/} arbitrary  (differentiable) 
functions of the variables $(v,x^A)$ .
Under such circumstances, the $\kappa$ in (\ref{ghoriz}) is a constant which coincides with the surface gravity of the Killing vector 
$K$ at the horizon
(this actually holds also for a wide class of non-vacuum metrics, satisfying an energy inequality $T_{\alpha\beta}K^\alpha K^\beta\geq 0$ at the horizon).
The horizon is extremal if and only if 
    \[ \kappa=0. \]
Then, the metric (\ref{ghoriz}) admits the (pointwise) limit 
	\[ \epsilon\ \rightarrow\ 0 \]
of the  transformation
\begin{equation}
\label{fepsilon} f_\epsilon: (v,u,x^A)\ \mapsto\ (\epsilon v,\frac{1}{\epsilon}u,x^A). 
\end{equation}
This is a generalization of the Horowitz limit mentioned before. The result is
\begin{equation}
{\rm lim}_{\epsilon\rightarrow 0} f_\epsilon^*g\ =\  
g_{AB}(0,x)dx^Adx^B - 2du\left(dv -2v\omega_A(0,x)dx^A + \frac{1}{2}v^2H(0,x))du\right).  
\end{equation}
As the limit of the transformation (\ref{fepsilon}), the resulting metric is $f_\epsilon$ invariant, that is, in addition to the original Killing vector field 
\begin{equation}
K=\partial_u
\end{equation}
it also has a second Killing vector field 
\begin{equation}
K^{(0)}\ =\ v\partial_v -u\partial_u . 
\end{equation}
Notice, that the surface $v=0$ is a Killing horizon also for $K^{(0)}$. It is, however, not an \emph{extremal\/} horizon with respect to this second Killing field.

Moreover, every surface $ u\ =\ u_0$ is the Killing horizon of a Killing vector field
	\[ K^{(0)}+u_0K. \] 

\section{Conclusions, outlook}

Another generalization of (\ref{theequation}, \ref{themetric}) is also possible. Consider
a spacetime foliated by non-expanding horizons. Let the foliation be defined by  
\begin{equation}
u\ =\ {\rm const} . 
\end{equation}
Then, a distinguished null vector field $\ell$ tangent to all the horizons is
\begin{equation}
\ell^\mu\ =\ g^{\mu\nu}u_{,\nu} . 
\end{equation}
On every spacial section $S$ of a  leaf 
\[  u\ =\ u_0 \]
of the foliation, the rotation 1-form potential $-\omega_A$ is defined by
\begin{equation}
\nabla_A\ell\ =\ -\omega_A\ell . 
\end{equation}
Suppose the spacetime satisfies vacuum Einstein's equations.  Then,
a constraint equation implied by Einstein's equations is again(!)
\begin{equation}
\label{theequation'} D_{(A}\omega_{B)}\ +\ \omega_A\omega_B\ -\ \frac{1}{2}R_{AB}\ =\ 0 
\end{equation}
(we use the same notation as in (\ref{theequation})). It has to be satisfied on every
leaf of the foliation independently (given a leaf, the independence of  choice of slice $S$ is
automatic, and follows from the $0$th low of non-expanding horizons). So, it takes the mathematical form of the extremal horizon constraint even though there is no visible extremal horizon in the spacetime (and perhaps none at all). This mystery is yet to be understood. 
 
Constructing  the family of solutions (\ref{themetric}) to Einstein's equations
by composing isolated horizons not only was a link between the theories of isolated
horizons and theories of near horizon geometries (surprisingly,  the paper \cite{PLJ} is unknown 
in the living reviews of those theories   \cite{LivRevIH,LivRevNHG}), but it also still contains  intriguing mysteries.
Throughout the paper we were assuming that the spacetime dimension is 4. Most of the reasoningsubsubsection{Inv}
generalizes to higher dimensions \cite{Reall,LPhigh,LivRevNHG
} and there is still potential
for new results. A non-axially symmetric solution to the equation (\ref{theequation}) would produce
a new solution to Einstein's equations. Understanding the mysterious emergence of the extremal horizon constraint in the context of the spacetimes composed from non-extremal isolated horizons could also prove interesting.

\section*{Acknowledgments}
We would like to thank Maciej Dunajski for pointing out the reference \cite{Reall}.
This work was partially supported by the Polish National
Science Centre grant No.~2015/17/B/ST2/02871.

\end{document}